\documentclass[aps,pra,showpacs]{revtex4}
\usepackage{graphicx}
\usepackage{dcolumn}
\usepackage{bm}

\begin{document}
\title{Accidental cloning of a single photon qubit in two-channel
continuous-variable quantum teleportation}
\author{Toshiki Ide}
\affiliation{%
Okayama Institute for Quantum Physics,\\
1-9-1 Kyoyama, Okayama City, Okayama, 700-0015, Japan\\
e-mail: toshiki\_ide@pref.okayama.jp}
\author{Holger F. Hofmann}%
\affiliation{%
JST-CREST, Graduate School of Advanced Sciences of Matter, Hiroshima
University,\\
Kagamiyama 1-3-1, Higashi Hiroshima 739-8530, Japan\\
e-mail: hofmann@hiroshima-u.ac.jp
}%

\begin{abstract}
The information encoded in the polarization of a single photon
can be transferred to a remote location by two-channel continuous
variable quantum teleportation. However, the finite entanglement
used in the teleportation causes random changes in photon number.
If more than one photon appears in the output, the continuous
variable teleportation accidentally produces clones of the original
input photon. In this paper, we derive the polarization statistics
of the $N$-photon output components and show that they can be
decomposed into an optimal cloning term and completely unpolarized
noise. We find that the accidental cloning of the input
photon is nearly optimal at experimentally feasible squeezing levels,
indicating that the loss of polarization information is
partially compensated by the availability of clones.
\end{abstract}


\maketitle

\section{Introduction}
One of the most fascinating aspects of quantum optics is the insight
it offers into the relation between the continuous field variables
and photon numbers. In many cases, quantum protocols can be implemented
by using either a photon or a continuous variable approach. In particular,
this is true for teleportation and cloning, where both approaches have
been realized experimentally \cite{Bou97,Fur98,Mar02,Lam02,Irv04,And05}.
Recently, there have also been efforts to combine both approaches,
e.g. by applying homodyne detection to photon number states
\cite{Lvo01,Lvo02},
or by adding and subtracting photons from squeezed and coherent light
\cite{Opa00,Zav04,Our06}. In the light of these technological
advances, it is
interesting to take a closer look at some of the possibilities inherent
in the application of continuous-variable protocols to photon number
states.

Since continuous-variable teleportation works for any input state,
it is in principle straightforward to apply it to photon number
inputs \cite{Pol99,Ral00,Ral01,Ide01}. However, the transmission
process does not
preserve photon number, so it is necessary to evaluate the effects
of photon loss and photon addition. Specifically, a qubit encoded
in the polarization state of a single photon can be either lost
or multiplied in the continuous-variable teleportation process.
If the photon is multiplied, the quantum information carried by its
polarization is distributed to all output photons, resulting in
an accidental cloning of the initial qubit. Photon multiplication
errors should therefore be evaluated in terms of their cloning
fidelity, which is of course limited by the fact that ideal cloning
of quantum states is impossible \cite{Woo96,Gis97}.

In the present paper, we analyze the photon number statistics in
the output of a qubit teleportation and show that it can be
decomposed into an optimal cloning term and completely unpolarized
white noise. We derive the cloning fidelities and show that
they are close to the optimal cloning fidelities at experimentally
feasible squeezing levels. This result indicates that the transfer of
quantum information is mostly limited by the availability of clones
in the output. Interestingly, photons are cloned even though the
transmitted field signal is not amplified. As our discussion shows,
the cloning effects can be quantified in terms of the Gaussian field
noise added in the teleportation process. We can therefore conjecture
that accidental cloning is a general effect of Gaussian field noise
on photonic qubits.

The rest of the paper is organized as follows.
In section \ref{sec:trans}, we formulate the transfer operator
for the continuous-variable teleportation of polarized light.
In section \ref{sec:qubit}, we apply the formalism to a single
photon state of unknown polarization and show that the output
density matrix can be separated into a mixture of optimal clones
and unpolarized white noise. In section \ref{sec:clone}, the
$N$-photon outputs are identified and the cloning fidelities
are derived. Finally, we summarize the results and their
possible relevance in section \ref{sec:conc}.

\section{Transfer of polarization by two-mode continuous-variable
teleportation}
\label{sec:trans}

Conventional continuous-variable teleportation transfers only a
single mode with a well defined polarization. In order to transfer
the polarization of a single photon, it is therefore necessary to
teleport two modes in parallel. Since continuous-variable teleportation
preserves the coherence of the modes, it is not important which
pair of orthogonal polarization modes is selected, as long as the
four mode entangled state used in the teleportation is unpolarized
\cite{Ide01,Dol03}.

\setlength{\unitlength}{1.2pt}
\begin{figure}[htbp]
\begin{picture}(200,155)

\put(80,20){\framebox(40,20){\large OPA}}

\put(80,42.5){\line(-1,1){22.5}}
\put(77.5,40){\line(-1,1){22.5}}
\put(55,65){\line(0,-1){5}}
\put(55,65){\line(1,0){5}}
\put(67.5,57.5){\makebox(10,10){\large $R$}}

\put(120,42.5){\line(1,1){22.5}}
\put(122.5,40){\line(1,1){22.5}}
\put(145,65){\line(0,-1){5}}
\put(145,65){\line(-1,0){5}}
\put(122.5,57.5){\makebox(10,10){\large $B$}}

\put(20,42.5){\line(1,1){22.5}}
\put(22.5,40){\line(1,1){22.5}}
\put(45,65){\line(0,-1){5}}
\put(45,65){\line(-1,0){5}}
\put(22.5,57.5){\makebox(10,10){\large $A$}}

\put(5,33){\makebox(20,6){Input}}
\put(5,23){\makebox(20,6){field}}

\put(50,55){\line(0,1){30}}
\put(40,45){\makebox(20,6){Beam}}
\put(40,35){\makebox(20,6){splitter}}

\put(45,77.5){\line(-1,1){22.5}}
\put(42.5,75){\line(-1,1){22.5}}
\put(20,100){\line(0,-1){5}}
\put(20,100){\line(1,0){5}}
\put(15,104.5){\makebox(10,10){\large $\vec{x}_-$}}

\put(55,77.5){\line(1,1){22.5}}
\put(57.5,75){\line(1,1){22.5}}
\put(80,100){\line(0,-1){5}}
\put(80,100){\line(-1,0){5}}
\put(75,104.5){\makebox(10,10){\large $\vec{y}_+$}}

\put(5,120){\framebox(90,35){}}
\put(30,140){\makebox(40,10){Measurement of}}
\put(30,125){\makebox(40,10){$\vec{\beta}=\vec{x}_- + i \vec{y}_+$}}

\bezier{200}(95,135)(125,130)(155,90)
\put(155,90){\line(0,1){10}}
\put(155,90){\line(-2,1){10}}

\put(137.5,68){\framebox(40,20){$\hat{D}(\vec{\beta})$}}

\put(170,92.5){\line(1,1){12.5}}
\put(172.5,90){\line(1,1){12.5}}
\put(185,105){\line(0,-1){5}}
\put(185,105){\line(-1,0){5}}

\put(175,120){\makebox(20,6){Output}}
\put(175,110){\makebox(20,6){field}}
\end{picture}
\caption{Schematic representation of the two-mode quantum teleportation
setup. The entangled state is generated by four-mode squeezing in an
optical parametric amplifier (OPA). Four separate homodyne detection
measurements are used to obtain the polarization components of the
complex displacement amplitudes $\vec{\beta}=(\beta_H , \beta_V)$, and
the corresponding complex two mode displacement amplitude is added to
the output field $B$.}
\label{setup}
\end{figure}
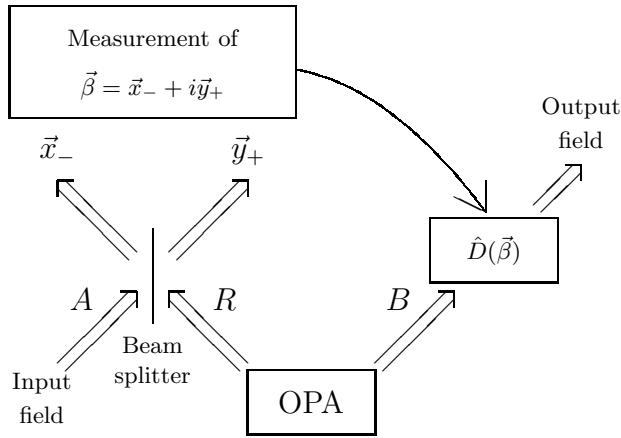
\setlength{\unitlength}{1pt}

Fig. \ref{setup} illustrates the extension of continuous-variable
teleportation to the two mode system using the Jones vectors of
the fields to express the polarization. The teleportation setup
uses the four mode squeezed state generated by an optical parametric
amplifier (OPA) to generate entanglement between two polarization
modes in beam $R$ and two polarization modes in beam $B$.
In the photon number basis of the horizontally ($H$) and vertically ($V$)
polarized modes, this four-mode squeezed entangled state can be written
as
\begin{equation}
\mid  \mbox{EPR}(q) \rangle_{R,B} = (1-q^2) \sum_{N=0}^{\infty}
q^{N} \sum_{n=0}^{N} \mid n; N-n\rangle_R \mid n; N-n\rangle_B.
\label{4modes}
\end{equation}
The amount of squeezing is given by $q$ ($0\leq q <1$), which is related
to the logarithmic attenuation of the field $r$ by $q=\tanh r$.
Since the amount of squeezing is the same for all polarizations,
the entangled state is completely unpolarized and the entanglement
between the polarizations of $R$ and $B$ in each $N$-photon subspace is
maximal.

The teleportation is then performed by mixing the two mode input in
beam $A$ with the reference $R$ at a 50/50 beam splitter and measuring
the quadrature components $\vec{x}_-$ of the difference and $\vec{y}_+$
of the sum. In practice, this requires four separate homodyne detection
measurements, to obtain the real and imaginary parts of the two
dimensional Jones vector, $\vec{\beta}=\vec{x}_- + i \vec{y}_+$.
However, the choice of polarizations for the Jones vector measurement
has no effect on the teleportation itself since the Bell measurement
performed on $R$ and $A$ simply projects the two beams onto a maximally
entangled state displaced by a coherent amplitude of $\vec{\beta}$
\cite{Hof00}.
Due to the entanglement between $R$ and $B$, the field value
$\vec{\beta}$ corresponds to a field difference between the unknown
input $A$ and the output beam $B$. This difference can be corrected
by a two mode field displacement $\hat{D}_{\mbox{\small pol}}
(\vec{\beta})$. The complete
teleportation process can then be summarized by a two mode transfer
operator, which is a straightforward extension of the single-mode
case derived in \cite{Hof00}. This operator reads
\begin{eqnarray}
\hat{T}_{\mbox{\small pol}}(\vec{\beta})&=& \frac{1-q^2}{\pi} \sum_{N=0}^
{\infty}
\sum_{n=0}^{N} q^{N}
\hat{D}_{\mbox{\small pol}}(\vec{\beta})
\hat{\Pi}_N
\hat{D}_{\mbox{\small pol}}(-\vec{\beta}),
\label{newtrans}
\end{eqnarray}
where the operator $\hat{\Pi}_N$ is the projection operator of the
$N$ photon subspace,
\begin{eqnarray}
\hat{\Pi}_N=\sum_{n=0}^N \mid n,N-n \rangle \langle n,N-n \mid.
\end{eqnarray}
It should be noted that the choice of polarization modes in the
preparation of the squeezed state and in the homodyne measurements
have no effect on the transfer process itself. The transfer operator
is completely defined by the Jones vector $\vec{\beta}$ obtained in
the homodyne detections.

As explained in \cite{Hof00}, the transfer operator determines both the
probabilities of measuring $\vec{\beta}$ and the output state of the
teleportation. The output density matrix is a mixture over all possible
measurement outcomes given by
\begin{eqnarray}
\hat{\rho}_{\mbox{\small out}} &=& \int d^4 \vec{\beta}\;
\hat{T}_{\mbox{\small pol}} (\vec{\beta})
\mid \psi_{\mbox{\small in}} \rangle \langle \psi_{\mbox{\small in}} \mid
\hat{T}_{\mbox{\small pol}}^{\dagger} (\vec{\beta}).
\label{trans}
\end{eqnarray}
In this integral, the Jones vector $\vec{\beta}$ is averaged over
all possible polarizations, so the polarization of the output depends
only on the input polarization.

Since it will be relevant in the following analysis, it may be useful
to consider the case of a vacuum input. In this case, the teleportation
simply adds Gaussian noise to the field, resulting in a thermal state
output given by
\begin{eqnarray}
\hat{R}_{\mbox{\small vac}}&=&
\int d^4 \vec{\beta}\;
\hat{T}_{\mbox{\small pol}} (\vec{\beta})
\mid 0;0 \rangle \langle 0;0 \mid
\hat{T}_{\mbox{\small pol}}^{\dagger} (\vec{\beta})
\nonumber \\
&=&
\left(\frac{1+q}{2}\right)^2 \sum_N \left(\frac{1-q}{2}
\right)^{N} \hat{\Pi}_N.
\label{R_vac1}
\end{eqnarray}
In the wave picture, the teleportation error can be interpreted as
Gaussian field noise with a variance of $V_q$ equal to the average
photon number added to each mode \cite{Bra98}. According to the thermal
state given by (\ref{R_vac1}), this error is related to the
squeezing parameter $q$ by
\begin{equation}
V_q=\frac{1-q}{1+q}.
\end{equation}
The value of the teleportation error $V_q$ lies between zero for
error free teleportation with infinitely squeezed light, and one
for the limit of classical teleportation using a pair of vacuum
states instead of entanglement. In the particle picture, $V_q$
is the average number of photons added per mode, so two-mode
teleportation adds a total average of $2 V_q$ photons in the
output field.

Since the vacuum state input is completely unpolarized, the
output state has no polarization either. However, the teleportation
process will transfer the polarization of the input state to the
output. In the next section, we will show how the transfer operator
can be used to describe this transfer of polarization in the
case of a single-photon polarization qubit.

\section{continuous-variable teleportation of a single-photon
polarization qubit}
\label{sec:qubit}

We now consider the case of a single photon input of unknown polarization.
Such input states can be described by a superposition of two basis
states, $\mid H \rangle = \hat{a}_H^\dagger \mid 0;0 \rangle$ and
$\mid V \rangle = \hat{a}_V^\dagger \mid 0;0 \rangle$. Alternatively,
the unknown quantum information can be described by a creation operator
$\hat{a}_{\mbox{\small in}}^{\dagger}$, so that the input state is given by
\begin{eqnarray}
\mid \psi_{\mbox{\small in}} \rangle &=&
\hat{a}_{\mbox{\small in}}^{\dagger} \mid 0;0 \rangle
\nonumber
\\[0.2cm]
\mbox{where}
\hspace{0.5cm}
\hat{a}_{\mbox{\small in}} &=& c_H^{\ast} \hat{a}_H + c_V^{\ast} \hat{a}_V.
\end{eqnarray}
By using the basis independent properties of the operator $\hat{a}_{\mbox
{\small in}}$, we can keep track of the quantum
information in the single photon input as it is transferred to
the multi photon output components.

According to the transfer operator formalism, the output density
matrix for the single photon qubit teleportation is given by
\begin{equation}
\hat{\rho}_{\mbox{\small out}}=\int d^4\vec{\beta}\; \hat{T}_{\mbox{\small
pol}}(\vec
{\beta})
\; \hat{a}_{\mbox{\small in}}^{\dagger} \mid 0; 0 \rangle
\langle 0; 0 \mid \hat{a}_{\mbox{\small in}} \hat{T}_{\mbox{\small pol}}^
{\dagger}(\vec{\beta}).
\label{rho_out1}
\end{equation}
To solve this integral, we need to consider the effects of the
transfer operator on the unknown operator $\hat{a}_{\mbox{\small in}}$,
which defines the polarization of the input qubit. For this
purpose, it is convenient to define the component $\beta_{\mbox{\small in}}$
of the Jones vector $\vec{\beta}$ with the same polarization
as the unknown input,
\begin{equation}
\beta_{\mbox{\small in}} = c_H^\ast \beta_H + c_V^\ast \beta_V.
\end{equation}
It is then possible to commute the transfer operators $\hat{T}_{\mbox{\small
pol}}(\vec
{\beta})$ and the operators of the unknown
input polarization, $\hat{a}_{\mbox{\small in}}$
and $\hat{a}_{\mbox{\small in}}^\dagger$, using the relations
\begin{eqnarray}
\hat{T}_{\mbox{\small pol}}(\vec{\beta}) \hat{a}_{\mbox{\small in}}^
{\dagger}
&=& (q \hat{a}_{\mbox{\small in}}^
{\dagger}+ (1-q)\beta_{\mbox{\small in}}^{\ast})
\hat{T}_{\mbox{\small pol}}(\vec{\beta})
\nonumber
\\
\hat{a}_{\mbox{\small in}}\hat{T}_{\mbox{\small pol}}(\vec{\beta})&=& \hat
{T}_{\mbox{\small pol}}(\vec
{\beta}) \left
( q \hat{a}_{\mbox{\small in}}+ (1-q)\beta_{\mbox{\small in}}\right).
\end{eqnarray}
The output density matrix can then be written as
\begin{eqnarray}
\hat{\rho}_{\mbox{\small out}}=\int d^4\vec{\beta}
\left(q^2 \hat{a}_{\mbox{\small in}}^{\dagger}
\hat{r}_{\mbox{\small vac}} \hat{a}_{\mbox{\small in}} +
(1-q)q ( \hat{a}_{\mbox{\small in}}^{\dagger}
\beta_{\mbox{\small in}} \hat{r}_{\mbox{\small vac}} +
\hat{r}_{\mbox{\small vac}} \beta_{\mbox{\small in}}^{\ast} \hat{a}_{\mbox
{\small in}})
+ (1-q)^2 \beta_{\mbox{\small in}} \hat{r}_{\mbox{\small vac}}
\beta_{\mbox{\small in}}^{\ast}
\right) ,
\label{rho_out2}
\end{eqnarray}
where $\hat{r}_{\mbox{\small vac}}$ is an abbreviation for the operator
obtained by applying the transfer operator to the vacuum density matrix,
\begin{eqnarray}
\hat{r}_{\mbox{\small vac}}(\vec{\beta})=\hat{T}_{\mbox{\small pol}}(\vec
{\beta}) \mid 0; 0 \rangle
\langle 0; 0 \mid \hat{T}_{\mbox{\small pol}}^{\dagger}(\vec{\beta}).
\end{eqnarray}
Since $\beta_{\mbox{\small in}}$ is unknown, the integral
in eq.(\ref{rho_out2}) still depends on the coefficients
$c_H$ and $c_V$ defining the direction of
the unknown input polarization. To obtain an integral that is independent
of the unknown polarization, we need to convert these values back into
operators independent of the measurement outcome $\vec{\beta}$.
This transformation can be accomplished by making use of the fact that
the vacuum teleportation $\hat{T}_{\mbox{\small pol}}(\vec{\beta})
\mid 0; 0 \rangle$ results in a coherent
state with an amplitude of $(1-q)\vec{\beta}$ \cite{Hof00}. Therefore,
this state is a right eigenstate of $\hat{a}_{\mbox{\small in}}$, and
we can transform $\beta_{\mbox{\small in}}$ into
$\hat{a}_{\mbox{\small in}}$ using
\begin{eqnarray}
\hat{a}_{\mbox{\small in}} \hat{r}_{\mbox{\small vac}}  &=&
(1-q) \beta_{\mbox{\small in}} \hat{r}_{\mbox{\small vac}}
\nonumber \\
\hat{r}_{\mbox{\small vac}} \hat{a}_{\mbox{\small in}}^{\dagger} &=&
(1-q) \beta_{\mbox{\small in}}^{\ast} \hat{r}_{\mbox{\small vac}}.
\end{eqnarray}
It is thus possible to convert all factors of $\beta_{\mbox{\small in}}$
in eq.(\ref{rho_out2}), resulting in an integral where only
$\hat{r}_{\mbox{\small vac}}$ depends on $\vec{\beta}$,
\begin{eqnarray}
\hat{\rho}_{out}=\int d^4\vec{\beta}
\left(q^2 \hat{a}_{\mbox{\small in}}^{\dagger}
\hat{r}_{\mbox{\small vac}} \hat{a}_{\mbox{\small in}} +
q \left(\hat{a}_{\mbox{\small in}}^{\dagger}
\hat{a}_{\mbox{\small in}} \hat{r}_{\mbox{\small vac}} +
\hat{r}_{\mbox{\small vac}} \hat{a}_{\mbox{\small in}}^{\dagger}
\hat{a}_{\mbox{\small in}} \right) +
\hat{a}_{\mbox{\small in}} \hat{r}_{\mbox{\small vac}}
\hat{a}_{\mbox{\small in}}^{\dagger} \right).
\label{rho_out3}
\end{eqnarray}
We can now perform the integration of $\hat{r}_{\mbox{\small vac}}$,
the result of which is equal to the output state of vacuum
teleportation $\hat{R}_{\mbox{\small vac}}$ given in
eq. (\ref{R_vac1}). The output of single photon teleportation can
therefore be expressed in terms of applications of the input
operator $\hat{a}_{\mbox{\small in}}$
to the unpolarized thermal state $\hat{R}_{\mbox{\small vac}}$,
\begin{eqnarray}
\hat{\rho}_{\mbox{\small out}}
&=& q^2 \hat{a}_{\mbox{\small in}}^{\dagger} \hat{R}_{\mbox{\small vac}}
\hat{a}_{\mbox{\small in}}
+ q \left( \hat{R}_{\mbox{\small vac}} \hat{a}_{\mbox{\small in}}^{\dagger}
\hat{a}_{\mbox{\small in}}
+ \hat{a}_{\mbox{\small in}}^{\dagger} \hat{a}_{\mbox{\small in}}
\hat{R}_{\mbox{\small vac}} \right)
+ \hat{a}_{\mbox{\small in}} \hat{R}_{\mbox{\small vac}}
\hat{a}_{\mbox{\small in}}^{\dagger}.
\label{rho_out4}
\end{eqnarray}
The result can be simplified by considering the commutation relations
of the thermal state and arbitrary creation and annihilation operators,
namely
\begin{eqnarray}
\hat{a}_{\mbox{\small in}} \hat{R}_{\mbox{\small vac}}
&=& \frac{1-q}{2} \hat{R}_{\mbox{\small vac}} \hat{a}_{\mbox{\small in}}
\nonumber \\
\hat{R}_{\mbox{\small vac}} \hat{a}_{\mbox{\small in}}^\dagger
&=& \frac{1-q}{2} \hat{a}_{\mbox{\small in}}^\dagger \hat{R}_{\mbox{\small
vac}}.
\end{eqnarray}
It is then possible to rearrange the operator ordering in
eq.(\ref{rho_out4}) so that the operator
$\hat{a}_{\mbox{\small in}}$ acts only as a creation operator
on the thermal output $\hat{R}_{\mbox{\small vac}}$. In this simplified
form,
the output reads
\begin{eqnarray}
\label{eq:separate}
\hat{\rho}_{\mbox{\small{out}}}
&=&
\left(\frac{1+q}{2}\right)^2
\underbrace{
\hat{a}_{\mbox{\small{in}}}^{\dagger} \hat{R}_{\mbox{\small vac}}
\hat{a}_{\mbox{\small{in}}}}
_{\mbox{photon added state}}
 + \left(\frac{1-q}{2}\right)
\underbrace{
\hat{R}_{\mbox{\small vac}}.
}_{\mbox{white noise}}
\label{rho2}
\end{eqnarray}
As indicated, it is now possible to interpret the output as
a mixture of a photon added state polarized by the application
of the creation operator $\hat{a}_{\mbox{\small in}}^\dagger$
to $\hat{R}_{\mbox{\small vac}}$, and a completely unpolarized
white noise component represented by the thermal state
$\hat{R}_{\mbox{\small vac}}$ itself.
It may be interesting to note that the photon added state is
a two mode version of the single photon added thermal state
investigated in recent experiments \cite{Par06,Zav07} because of
its non-classical features such as the negativity of the Wigner
function \cite{Aga92}. The photon added term in eq.(\ref{eq:separate})
thus describes the teleportation of non-classical features in the
field quadrature statistics of the single photon state.
For the purpose of photon cloning however, only the photon number distributions are relevant. In that context, it is significant that
the application of a creation operator also describes the effects
of photon bunching and
of stimulated emission used in previous photon cloning experiments
\cite{Lam02,Mar02,Irv04}. As we shall show in the following, the
photon added state is indeed equivalent to a mixture of
optimally cloned $N$-photon outputs. It is therefore possible to interpret the photon added state as optimal cloning and the
thermal white noise background as a measure of the non-optimal
nature of accidental cloning.

More specifically, eq.(\ref{eq:separate}) provides a simple
quantification of the unpolarized white noise background added in
the teleportation in terms of the statistical weights of the two
components. Since $\hat{R}_{\mbox{\small vac}}$ is normalized, the statistical weights are $(1-q)/2$ for the white noise and $(1+q)/2$
for the photon added state representing optimal cloning. The ratio
of the two components is thus exactly equal to the Gaussian field
error $V_q$ derived in sec. \ref{sec:trans}. Since $V_q$ is a
very intuitive measure of the teleportation error in terms of
field noise, it may be interesting to see how this continuous
variable noise measure is related to the cloning errors in the
discrete photon number statistics.

\section{Cloning fidelity of the $N$-photon output}
\label{sec:clone}

In the previous section, we have derived the complete output state
of the two mode teleportation of a single photon polarization qubit.
The total output photon number of this state is random, so it
is not possible to predict how many clones (if any) are generated.
Since the optimal cloning fidelity is a function of the number of
clones produced, it is necessary to separate the output density
matrix of eq.(\ref{rho2}) into its $N$-photon output components,
representing the accidental occurrences of $1 \to N$ cloning
events,
\begin{equation}
\hat{\rho}_{\mbox{\small out}} =
\sum_{N=0}^{\infty} P(N) \hat{\rho}_N.
\end{equation}
The decomposition of the unpolarized thermal states
$\hat{R}_{\mbox{\small vac}}$ in eq.(\ref{rho2}) is given by
eq.(\ref{R_vac1}). In the photon added component that represents
optimal cloning, the number
of photons is raised by one due to the application of the creation
operator $\hat{a}_{\mbox{\small in}}^\dagger$. Therefore, the
decomposition of the output density matrix into $N$-photon subspaces
includes a white noise term given by $\hat{\Pi}_N$ and an optimal
cloning term given by
$\hat{a}_{\mbox{\small{in}}}^{\dagger} \hat{\Pi}_{N-1}
\hat{a}_{\mbox{\small{in}}}$. Since the optimal cloning part
has no zero-photon component, $\hat{\Pi}_{-1}$ should be defined
as zero. The decomposition of the output density matrix into $N$-photon
subspaces then reads
\begin{eqnarray}
\label{eq:Nderive}
\hat{\rho}_{\mbox{\small out}} &=&
\frac{(1+q)^2}{2(1-q)}\sum_{N=0}^{\infty}
\left( \frac{1-q}{2} \right)^N
\left( \left( \frac{1+q}{2} \right)^2 \hat{a}_{\mbox{\small{in}}}^{\dagger}
\hat{\Pi}_{N-1} \hat{a}_{\mbox{\small{in}}} + \left( \frac{1-q}{2} \right)^2
\hat{\Pi}_N \right) \nonumber \\
&=&
\frac{1}{V_q (1+V_q)^3}
\sum_{N=0}^{\infty}
\left( \frac{V_q}{1+V_q} \right)^N
\left( \hat{a}_{\mbox{\small{in}}}^{\dagger} \hat{\Pi}_{N-1} \hat{a}_{\mbox
{\small{in}}} + V_q^2 \hat{\Pi}_N \right).
\label{rho3}
\end{eqnarray}
Here, the entanglement parameter $q$ has been converted into the
more intuitive measure of Gaussian field error $V_q$. It is then possible
to relate the photon number distribution and the cloning errors directly
to the Gaussian noise error of the continuous variable teleportation.

The statistical weights of the contributions in eq.({\ref{eq:Nderive})
are determined by the traces of the operators. Specifically, the
trace of the white noise term $\hat{\Pi}_{N}$ is $N+1$ and the
trace of the optimal cloning term
$\hat{a}_{\mbox{\small{in}}}^{\dagger} \hat{\Pi}_{N-1} \hat{a}_
{\mbox{\small{in}}}$ is $(N+1)N/2$.
The probability $P(N)$ of obtaining an $N$-photon output is
given by the product trace of the total density matrix
in the $N$-photon subspace. In terms of the Gaussian field
noise $V_q$, this photon number distribution reads
\begin{eqnarray}
P(N) &=& \mbox{Tr}\{\hat{\Pi}_N  \hat{\rho}_{\mbox{\small out}}\}
\nonumber \\[0.2cm]
&=& \frac{(N+1)(N+2 V_q^2)}{2 V_q (1+V_q)^3}\;
\left(\frac{V_q}{1+V_q} \right)^N.
\end{eqnarray}
$P(0)$ is the probability of losing the photon, $P(1)$ is
the probability of single photon teleportation, and
$P(N \geq 2)$ are the probabilities of accidental
$1 \to N$ cloning.
Since the generation of additional photons is
itself a kind of teleportation error, the probability of
generating accidental clones increases with $V_q$.
Specifically, the average photon number in the output is
$1+2 V_q$ (the original photon plus twice the average photon
number added per mode). Even at $V_q=1$, the average photon
number is only three and the probability of obtaining a high
number of output photons drops with $(1/2)^N$. Accidental
cloning probabilities are therefore generally low for high
numbers of clones. However, the probabilities of obtaining
two, three or four clones can be quite significant, as shown in
fig.\ref{P(N)}. Specifically, the probabilities of obtaining
$N$ clones at an experimentally feasible error of $V_q=0.25$
requiring about 6 dB squeezing are $P(2)=26.1 \%$, $P(3)=10.2 \%$, and
$P(4)= 3.4 \%$. The accidental generation of clones should therefore
be a very common occurrence if qubits are teleported using squeezed
state entanglement at presently available levels of squeezing.

\begin{figure}[htbp]
\begin{picture}(260,160)
\put(220,0){\makebox(40,10){$V_q$}}
\put(80,115){\makebox(40,10){$N=2$}}
\put(80,85){\makebox(40,10){$N=3$}}
\put(110,60){\makebox(40,10){$N=4$}}
\put(0,145){\makebox(40,10){$P(N)$}}
\put(72,150){\line(0,-1){5}}
\put(72,140){\line(0,-1){5}}
\put(72,130){\line(0,-1){5}}
\put(72,120){\line(0,-1){5}}
\put(72,110){\line(0,-1){5}}
\put(72,100){\line(0,-1){5}}
\put(72,90){\line(0,-1){5}}
\put(72,80){\line(0,-1){5}}
\put(72,70){\line(0,-1){5}}
\put(72,60){\line(0,-1){5}}
\put(72,50){\line(0,-1){5}}
\put(72,40){\line(0,-1){5}}
\put(72,30){\line(0,-1){5}}
\put(72,20){\line(0,-1){5}}
\put(72,10){\line(0,-1){5}}
\includegraphics[width=8cm]{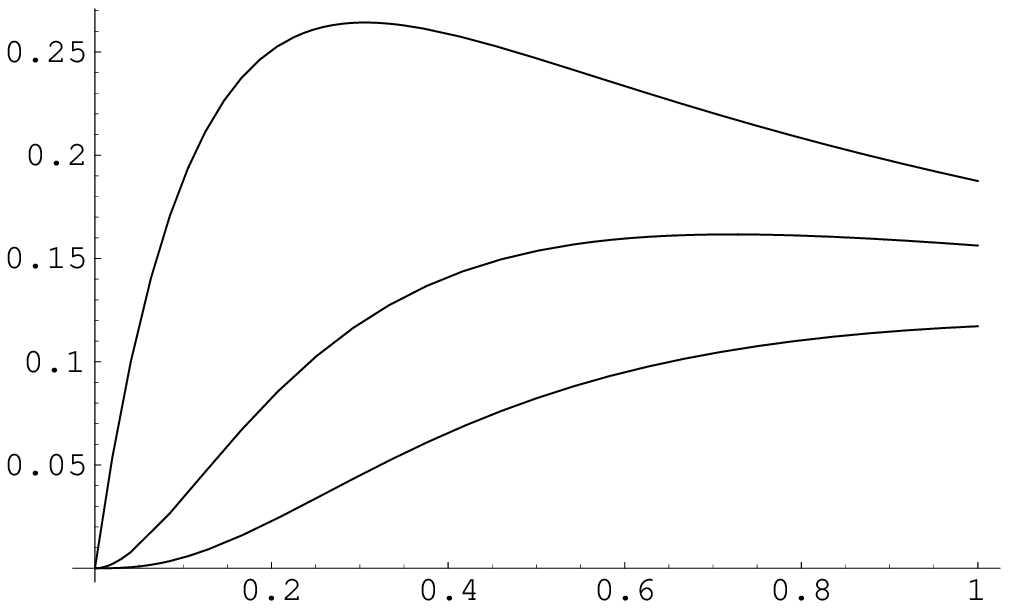}
\end{picture}
\
\caption{Cloning probabilities $P(N)$ for different output photon
numbers $N$ plotted to $V_q$. From top to bottom, the curves show
$P(2), P(3), P(4)$. The dashed vertical line indicates $V_q=0.25$.}
\label{P(N)}
\end{figure}

We can now evaluate the quality of accidental $1 \to N$ cloning
by separating the $N$-photon output $\hat{\rho}_N$ into its optimal
cloning component $\hat{C}_N$ and its white noise component $\hat{W}_N$,
\begin{equation}
\hat{\rho}_N = \eta_N \hat{C}_N + (1-\eta_N) \hat{W}_N.
\end{equation}
The density matrices $\hat{C}_N$ and $\hat{W}_N$ are independent
of the teleportation errors $V_q$ and can be generated from
the projection operators $\hat{\Pi}_N$ into the $N$-photon subspaces by
\begin{eqnarray}
\hat{C}_N
&=&
\frac{2}{N(N+1)}
\hat{a}_{\mbox{\small in}}^{\dagger}\hat{\Pi}_{N-1}\hat{a}_{\mbox{\small
in}}\nonumber \\[0.3cm]
\hat{W}_N
&=&
\frac{1}{N+1}\hat{\Pi}_N.
\label{pcwn}
\end{eqnarray}
Therefore, the $N$-photon output is fully characterized by the single
parameter $\eta_N$. Since this parameter defines the fraction of
optimally cloned $N$-photon outputs, we will call it the cloning
efficiency of accidental $1 \to N$ cloning. The cloning efficiency
$\eta_N$ is a function of teleportation error $V_q$ and photon
number $N$. Using eq.(\ref{rho2}), we find that this ratio is
\begin{equation}
\eta_N = \frac{1}{1+2 V_q^2/N}.
\label{etacn}
\end{equation}

Fig. \ref{eta} illustrates this dependence of cloning efficiency $\eta_N$ on the teleportation error $V_q$ for several output photon
numbers $N$. Not surprisingly, the teleportation error $V_q$ reduces
the cloning efficiency. However, it is interesting to note that
only the square
of $V_q$ enters into the relation, indicating that the cloning
efficiency rapidly approaches one for low teleportation errors.
For example, an experimentally feasible error of $V_q=0.25$ requiring
about 6 dB of squeezing already gives a two photon cloning efficiency
of $\eta_2=16/17$, or about 94 \%. We can therefore conclude that
the accidental cloning observed in continuous variable teleportation
at presently available levels of squeezed state entanglement
will be nearly optimal.
Another significant feature of accidental cloning is
that the minimal cloning efficiencies obtained at the classical teleportation limit ($V_q=1$) have a
photon number dependent value of $N/(N+2)$. Thus, the cloning
efficiencies for high $N$ are always close to one, indicating that
the generation of a large number of clones is quite robust
against teleportation errors.

\begin{figure}[htbp]
\begin{picture}(260,160)
\put(210,135){\makebox(40,10){$V_q$}}
\put(220,110){\makebox(40,10){$N=100$}}
\put(220,45){\makebox(40,10){$N=4$}}
\put(220,22){\makebox(40,10){$N=3$}}
\put(170,2){\makebox(40,10){$N=2$}}
\put(0,130){\makebox(10,10){$\eta_N$}}
\includegraphics[width=8cm]{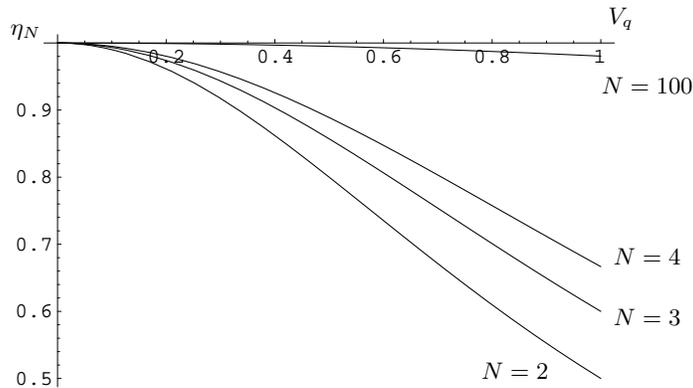}
\end{picture}
\caption{Cloning efficiency $\eta_N$ for different output photon
numbers $N$ plotted to $V_q$. From bottom to top, the curves show
$\eta_2, \eta_3, \eta_4$ and $\eta_{100}$.}
\label{eta}
\end{figure}

Up to now, our discussion was based only on the formal decomposition
of the output into an optimal cloning component and a white noise
term. Experimentally, this decomposition is not directly observable.
Instead, cloning is characterized by the cloning fidelity $F_N$,
defined as the fraction of output photons with the same polarization
as the input photon, $\langle \hat{a}_{\mbox{\small in}}^\dagger \hat{a}_{\mbox{\small in}} \rangle/N$.
This fidelity is usually measured by splitting up
the output light into a sufficiently large number of channels,
so that the polarization of each photon can be detected separately
\cite{Lam02,Irv04}. It is then possible to evaluate cloning by
conventional photon counting.

To determine the fidelity $F_N$ of a $1\to N$ cloning process with
cloning efficiency $\eta_N$, we can use the fidelities of the optimal
cloning component $\hat{C}_N$ and the white noise components
$\hat{W}_N$. Since the white noise component is completely unpolarized,
exactly half of the photons will have the same polarization as the
input, corresponding to a fidelity of $1/2$. For the optimal cloning
term, the cloning fidelity is given by
\begin{equation}
\label{eq:opt}
F_{\mbox{\small opt.}} =
\frac{1}{N} \mbox{Tr}\left\{
\hat{a}_{\mbox{\small in}}^{\dagger}\hat{a}_{\mbox{\small in}}
\hat{C}_N
\right\} = \frac{2N + 1}{3N}.
\end{equation}
This result is equal to the $1\to N$ cloning fidelity of an
optimal cloning machine \cite{Gis97}, proving our conjecture that
$\hat{C}_N$ represents optimal cloning.
The accidental cloning fidelity $F_N$ of the mixture of $\hat{C}_N$ and
$\hat{W}_N$ defined by the cloning efficiency $\eta_N$
can now be obtained by taking the weighted
average of the white noise fidelity of $1/2$ and the optimal
cloning fidelity given by eq.(\ref{eq:opt}),
\begin{eqnarray}
F_N
&=&
\eta_{N} \frac{2N+1}{3N} + \left(1-\eta_{N}\right) \frac{1}{2}
\nonumber \\
&=&
\frac{2}{3} + \frac{1-V_q^2}{3 (N+2 V_q^2)}.
\label{cloning}
\end{eqnarray}
As this relation shows, the cloning efficiencies at $V_q=1$ all
correspond to a cloning fidelity of $2/3$, which is the optimal
cloning fidelity for $N \to \infty$. Moreover, $2/3$ is also the
optimal cloning fidelity obtained if the cloning is performed
using only local measurements and classical communications instead
of a direct quantum mechanical interaction between the original and
the clones. At $V_q=1$, this is exactly what happens, since there
is no entanglement and the teleportation is performed by a local
measurement of the input photon and the generation of a coherent
state of the appropriate amplitude in the output. It is therefore
obvious that $V_q=1$ is still close to the optimal limit for
high $N$. On the other hand, cloning fidelities above $2/3$ are
only possible because of the entanglement used in the teleportation.

\begin{figure}[htbp]
\begin{picture}(260,160)
\put(220,20){\makebox(40,10){$V_q$}}
\put(100,95){\makebox(40,10){$N=2$}}
\put(40,85){\makebox(40,10){$N=3$}}
\put(40,45){\makebox(40,10){$N=4$}}
\put(100,7){\makebox(40,10){$N=100$}}
\put(10,135){\makebox(40,10){$F_{N}$}}
\put(76,130){\line(0,-1){5}}
\put(76,120){\line(0,-1){5}}
\put(76,110){\line(0,-1){5}}
\put(76,100){\line(0,-1){5}}
\put(76,90){\line(0,-1){5}}
\put(76,80){\line(0,-1){5}}
\put(76,70){\line(0,-1){5}}
\put(76,60){\line(0,-1){5}}
\put(76,50){\line(0,-1){5}}
\put(76,40){\line(0,-1){5}}
\put(76,30){\line(0,-1){5}}
\put(76,20){\line(0,-1){5}}
\put(76,10){\line(0,-1){5}}
\put(76,0){\line(0,-1){5}}
\includegraphics[width=8cm]{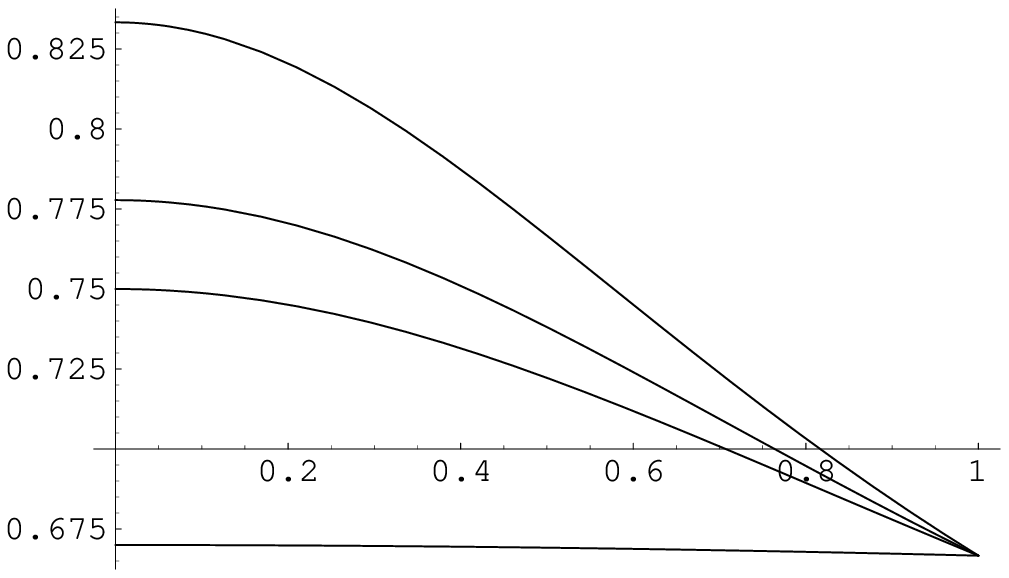}
\end{picture}
\caption{Cloning fidelities $F_{N}$ for different output photon
numbers $N$ plotted to $V_q$. From top to bottom, the curves show
$F_{2}, F_{3}, F_{4}$ and $F_{100}$. The dashed vertical line indicates
$V_q=0.25$.}
\label{cfid2}
\end{figure}

Fig.\ref{cfid2} shows the cloning fidelity $F_N$ for two, three,
and five output photons. In addition, the high $N$ limit is indicated
by the result for $N=100$, which is basically indistinguishable from
a flat line at $F_{\infty} =2/3$. The dotted line indicates
a realistic teleportation error of $V_q=0.25$, achievable by
about 6 dB of squeezing. The cloning fidelities at this noise level
are already quite close to the optimal fidelities. Specifically,
the fidelities at $V_q=0.25$ ($V_q=0$) are
$F_2=0.814$ ($F_2=0.833$), $F_3=0.767$ ($F_3=0.778$),
$F_4=0.742$ ($F_4=0.75$). It should therefore be possible to observe
nearly optimal accidental quantum cloning in the continuous
variable teleportation of single photon qubits under presently
realizable experimental conditions.

In order to put the above results into a wider context, it may be
interesting to recall that the cloning effect has been obtained
without any field amplification. In fact, a minimal noise
amplification can produce optimal clones, and this situation can
be realized by adjusting the gain of continuous variable teleportation,
as we have already pointed out elsewhere \cite{Hof05}. However,
accidental cloning occurs at a gain of one, due to the Gaussian field
noise added in the teleportation. Since all linear optics processes
should be equivalent, we can conjecture that our results describe
the accidental cloning effects of any kind of Gaussian field noise
added to the field of a single photon qubit state. In particular,
$V_q$ can be modified to include a variety of Gaussian errors in
addition to the limitation of squeezed state entanglement. In this
case, $V_q$ could even exceed one, indicating fidelities below $2/3$.
However, all successful continuous variable teleportation experiments
should reduce $V_q$ well below one, corresponding to cloning fidelities
close to optimal cloning.

\section{Conclusions}
\label{sec:conc}

Our results show that the accidental multiplication of a single
photon polarization qubit transferred by continuous variable
teleportation can be interpreted as a nearly optimal quantum
cloning process. A continuous-variable teleportation system operating at
feasible squeezing levels thus tends to act both as a fax and a copy
machine on the teleported qubits.
Even though the cloning happens as a consequence of the noise
introduced by non-maximal entanglement, the effect is similar
to intentional telecloning, where a special multi-photon entangled
state needs to be prepared beforehand \cite{Mur99}. Our analysis
indicates that telecloning is a natural feature of the
continuous-variable teleportation process when it is applied
to single-photon qubits. This observation might be useful in
the implementation of multi party protocols, where the distribution
of quantum information to several parties is desirable.

It is also interesting to note that the clones are generated without
field amplification, simply by the random addition of Gaussian field
noise. We can therefore conjecture that accidental cloning is a general
effect of Gaussian field noise on photonic qubits.
Even though this kind of cloning without amplification can never be
optimal, the additional error can be described by mixing the optimal
cloning output with a completely unpolarized component of the density
matrix. We have thus successfully converted a Gaussian field error into
an $N$-photon cloning error, with the cloning efficiency $\eta_N$
describing the exact fraction of the optimal cloning state in the
output. In the context of recent investigations into the quantum
mechanics of the photon-field dualism
\cite{Lvo01,Lvo02,Opa00,Zav04,Our06}, this result may shed some
light on the fundamental relations between photons and field noise.

In conclusion, the accidental cloning of photonic qubits in continuous
variable teleportation is a phenomenon that may be both useful in the
development of new technologies for quantum information networks and
in the exploration of the fundamental physics behind the dualism
of photons and fields. Hopefully, the analysis presented above will
be a fruitful contribution to both.

\section*{Acknowledgment}
Part of this work has been supported by the Grant-in-Aid program
of the Japanese Society for the Advancement of Science and by the
JST-CREST project on quantum information processing.

\end{document}